\documentclass[conference]{IEEEtran}
\IEEEoverridecommandlockouts

\usepackage{cite}
\usepackage{amsmath,amssymb,amsfonts}
\usepackage{algorithmic}
\usepackage{graphicx}
\usepackage{textcomp}
\usepackage{xcolor}
\def\BibTeX{{\rm B\kern-.05em{\sc i\kern-.025em b}\kern-.08em
    T\kern-.1667em\lower.7ex\hbox{E}\kern-.125emX}}

\usepackage{hyperref}
\usepackage{url}
\usepackage{float}
\usepackage[noabbrev,capitalize]{cleveref}

\usepackage{comment}
\usepackage{subcaption}
\usepackage{multirow}
\usepackage{multicol}
\usepackage{times}
\usepackage{array}
\usepackage{latexsym}
\usepackage{tabularx}
\usepackage{inconsolata}
\usepackage{enumitem}
\usepackage{tcolorbox}
\usepackage{subcaption}
\usepackage{ragged2e} 
\usepackage{listings}
\usepackage{placeins}
\usepackage{cuted}        
\usepackage{capt-of}      
\usepackage{booktabs}
\usepackage{makecell} 
\usepackage{dblfloatfix}
\usepackage{cuted}
\allowdisplaybreaks
\begin{document}

\title{Sparse Weight and Edge Circuit Discovery in Transformer-based Acoustic Models\\

\thanks{This work was supported by the Natural Sciences and Engineering Research Council of Canada (NSERC) ALLRP 588144, as well as resources provided by Compute Ontario and the Digital Research Alliance of Canada.}
}

\author{\IEEEauthorblockN{
Jiankun Wei}
\IEEEauthorblockA{\textit{Department of Computer Science} \\
\textit{University of Toronto}\\
Toronto, Canada \\
jiankun.wei@mail.utoronto.ca}
\and
\IEEEauthorblockN{ Ewan Dunbar}
\IEEEauthorblockA{\textit{Department of Computer Science} \\
\textit{University of Toronto}\\
Toronto, Canada \\
ewan.dunbar@utoronto.ca}
\and
\IEEEauthorblockN{ Gerald Penn}
\IEEEauthorblockA{\textit{Department of Computer Science} \\
\textit{University of Toronto}\\
Toronto, Canada \\
gpenn@cs.toronto.edu}

}

\maketitle

\begin{abstract}
Transformer-based foundation models are powerful but opaque, motivating Mechanistic Interpretation methods to uncover the black-box by identifying small computation subgraphs responsible for a task. DiscoGP is a joint weight-and-edge circuit discovery framework originally developed for text decoders. We extend DiscoGP to speech encoders and present, to our knowledge, the first circuit discovery study for modern speech foundation models. Across HuBERT and Wav2Vec 2.0 on several speech classification tasks, we find that the discovered circuits are extremely compact, yet often match or even exceed the performance of the full pretrained encoder with the same downstream head. Through ablations, we show that these circuits reflect pretrained computation rather than random structure or task-head artifacts. We also introduce a memory-efficient DiscoGP variant that reduces the GPU memory cost of edge-circuit discovery at runtime from quartic to cubic. Overall, our results broaden Mechanistic Interpretation beyond text decoders and show that circuit-level analysis can reveal both explanatory structure and unexpected functional behavior in speech encoders.

\end{abstract}


\section{Introduction}

Transformer-based foundation models~\cite{vaswani2017attention, devlin2019bert, 2020t5, touvron2023llama2, radford2019gpt2, openai2023gpt4} have become central to modern language and speech systems, yet they remain difficult to interpret or modify in a targeted way. Mechanistic interpretation seeks to address this by mapping model behavior to concrete computations inside the network. Circuit discovery is a particularly useful approach: instead of asking whether individual neurons or layers are meaningful in isolation, it aims to identify small task-relevant computation subgraphs whose collective behavior explains a model’s output. Early work relied on manual activation patching and analysis~\cite{wang2022interpretability}; later methods such as ACDC automated circuit search in GPT-2 and showed that very small edge sets can recover known behaviors~\cite{conmy2023automatedcircuit}. More recent work has extended circuit discovery with gradient-based methods and modular circuit composition~\cite{mondorf2025circuitcompositions, syed2024attributionpatching, hanna2024faithfulness, zhang2025eapgp}. DiscoGP~\cite{yu-etal-2025-sheaf} advances this line of work by jointly learning weight masks and edge masks, producing isolated circuits that preserve task behavior.


\begin{figure*}[t!]
    \begin{subfigure}[t]{0.5\textwidth}
    \centering
    \includegraphics[width=0.75\linewidth]{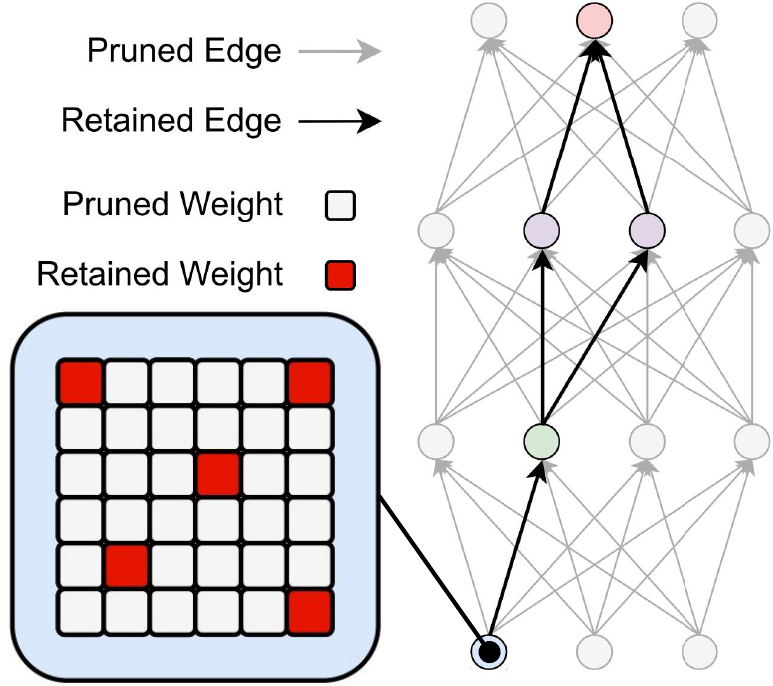}
    \caption{Overview of DiscoGP}
    \label{fig:figure1b}
    \end{subfigure}
    \begin{subfigure}[b]{0.5\textwidth}
    \includegraphics[width=\linewidth]{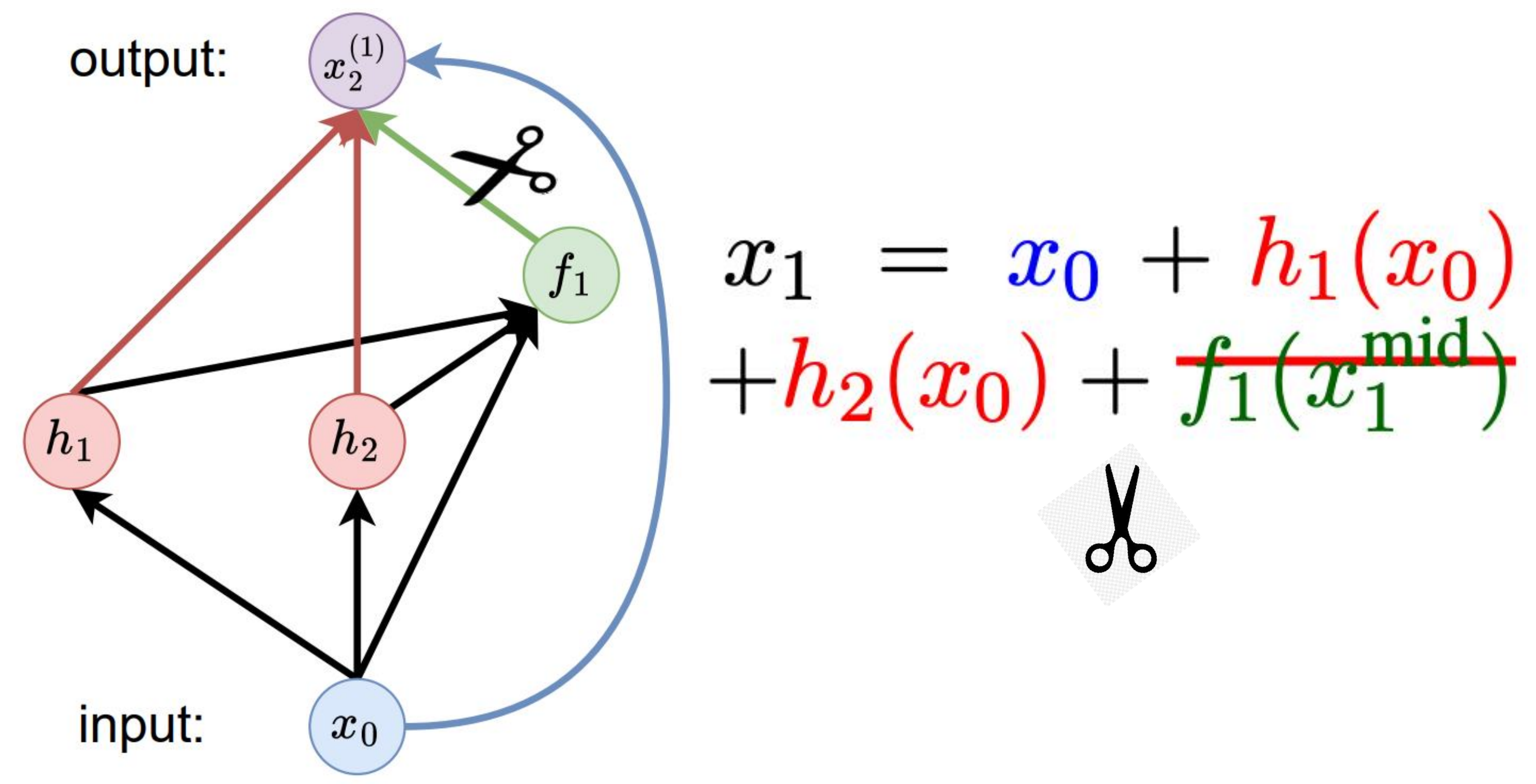}
    \caption{Discovering edge circuits}
    \label{fig:figure1c}
    \end{subfigure}
    \caption{(a) DiscoGP is the first mechanistic interpretation algorithm that combines the discovery of weight circuits and edge circuits. In this paper we extend DiscoGP to Encoder architectures and speech language models. (b) Edge circuits are subgraphs of the directed computation graph of a model. In this example, the MLP edge is cut to form a subgraph.}
    \label{fig:figure1}
    \vspace{-0.2in}
\end{figure*}
Despite these advances, circuit discovery has largely been studied in decoder-only text models, leaving encoder-style architectures and speech foundation models unexplored. This matters for two reasons. First, encoder representations are not directly exposed through autoregressive logits, so circuit discovery requires an auxiliary task head and a careful check that the discovered structure is not an artifact of that head. Second, no prior work has demonstrated circuit discovery in modern speech encoders at scale. We address this gap by adapting Circuit Discovery to HuBERT~\cite{hsu2021hubert} and Wav2Vec 2.0~\cite{baevski2020wav2vec2} and evaluating the discovered circuits on vowel, consonant, and voicing classification, keyword spotting, emotion recognition, and spoof detection across multiple datasets.
We emphasize that this work is about circuit discovery for mechanistic interpretability, not model compression. While DiscoGP often yields sparse subgraphs, our focus is on whether these subgraphs faithfully recover task-relevant computation in pretrained speech encoders, rather than on reducing parameters or deployment cost.

Our results show that speech encoders also admit tiny, task-specific circuits similar to their text decoder counterparts. Across tasks, DiscoGP consistently finds sparse subgraphs that preserve performance and in many cases outperform the full pretrained encoder with the same downstream head. Ablations with random and frozen models show that these circuits are not accidental artifacts of the architecture or head, but reflect meaningful pretrained computation. 

In summary, this work makes two contributions. It is, to our knowledge, the first large-scale circuit discovery study for speech encoders, and it provides a memory-efficient variant of DiscoGP that enables circuit search under tighter hardware constraints. Together, these results broaden mechanistic interpretability beyond text decoders and show that speech models contain compact, task-relevant circuits with both explanatory and practical value.
\section{Background}
The term \textit{circuit} is broadly interpreted as a subnetwork within a neural model, but its precise definition varies across the literature. Neel Nanda~\cite{nanda2023glossary} characterizes a circuit as “the sub-part of a model that performs some understandable computation to produce interpretable features from prior interpretable features.” Building on this intuition, several studies~\cite{olsson2022incontext,deepseek2024v3} have observed that the QK and OV matrices are always multiplied together in attention computations, allowing them to be treated as composite matrices. Based on this insight, Olsson et al.~\cite{olsson2022incontext} define the resulting composed QK and OV matrices as query–key and output–value circuits, respectively.

Following these developments, Wang et al.\cite{wang2022interpretability} treat individual attention heads within a transformer block as circuits and attempt to manually identify task-relevant circuits through \textbf{activation patching}: they corrupt the output of a single head with a counterfactual value and measure the resulting change in model output.
More recently, Conmy et al.\cite{conmy2023automatedcircuit} represent the computations of a transformer layer as a directed acyclic graph (computational graph) and search for subgraphs responsible for a specific behavior, enabling an automated form of circuit discovery that naturally leads to edge ciruits.

\subsection{Directed Acyclic Graph and Edge Circuit Discovery}

A transformer block consists of an attention module and an MLP module, and multiple blocks are stacked to form the encoder or decoder of a Transformer network. Let $x_{i-1}$ be the output of the (i-1)st transformer block (with $x_0$ denoting the input), let $H_{i}$ be the set of attention heads in block i, and $f_i$, its MLP layer. The output $x_i$ is defined as:
\[\begin{cases}
& x_{i}^{mid} = x_{i-1} + \sum_{h \in H_i} h(x_{i-1}) \\
& x_i = x_{i}^{mid} + f_i (x_{i}^{mid})
\end{cases}\]
Treating each term as an edge, and noting that each head computes $Q=W^Qx_{i-1}, K=W^Kx_{i-1}, V=W^Vx_{i-1}$, we obtain $3|H_i| + 2$ direct edges for $i = 1$. For deeper layers, each occurrence of $x_{i-1}$ can be recursively unrolled back to the initial input $x_0$, resulting in a quadratic number of indirect edges.

Edge circuit discovery aims to reduce this combinatorial explosion by identifying the smallest set of edges that forms a functional information-flow path from the input, through the stacked transformer blocks, to the final output by learning a binary mask over both direct and indirect edges.

\subsection{Weight Circuit Discovery} 
At the same time, a substantial body of work in interpretaton research focuses on \textbf{weight circuits discovery}: methods that single out compact subnetworks (specific subsets of a model’s parameters) that are responsible for particular functions or tasks ~\cite{cao2021lowcomplexity, csordas2021modular, decao2022sparseinterventions, zhang2021subnetwork, guo2021diffpruning}. This research trajectory was originally motivated by the  Lottery Ticket Hypothesis~\cite{frankle2019lotteryticket}, which posits that dense, randomly initialized neural networks contain ``winning ticket'' subnetworks that, when executed alone, can reach test accuracy comparable to the original network. However, weight circuits can equally be found in pre-trained models, allowing us to identify task-specific circuits created during training:  \cite{lepori2023breakitdown} adapted ideas of weight circuits to transformer-based language models and developed procedures such as continuous sparsification that can extract functional subnetworks directly from a trained model, preserving task performance without fully retraining~\cite{savarese2020continuoussparsification}.

\subsection{DiscoGP}
DiscoGP discovers task-specific circuits by jointly discovering both model \emph{weights} and \emph{edges} circuits.  Let \(G = (V,E)\) denote the model's computation graph and \(\theta\), its parameters.  DiscoGP learns a pair of binary masks:
\[
m = (m_{\theta}, m_{E}) \in \{0,1\}^{|\theta| + |E|}
\]
that gate weight parameters and edges; these masks are optimized by gradient-based learning.

\textbf{Gradient-based binary masks.}
Following prior work on gradient-based mask learning ~\cite{cao2021lowcomplexity, csordas2021modular, decao2022sparseinterventions, bayazit-etal-2024-discovering, louizos2018l0} each mask entry $m_i$ is modeled as a random variable from a Gumbel--sigmoid distribution parameterized by a learnable logit $l_i$. DiscoGP first computes a continuous score \(s_i\in[0,1]\) by injecting Gumbel noise and applying a temperatured sigmoid:
$$\begin{cases}
s_i &= \sigma\Big(\frac{l_i + \alpha_i - \beta_i}{\tau}\Big)\\
\alpha_i, \beta_i &\overset{\text{i.i.d.}}{\sim} \log(\log U)\\
U&\sim\mathrm{Uniform}(0,1)
\end{cases}$$

where \(\tau>0\) is the temperature and \(\sigma(\cdot)\) denotes the logistic sigmoid. 

\textbf{Sheaf search objectives.}
Given a task dataset $\mathcal{D}=\{(x_i,\hat{y}_i)\}$ where $\hat{y}_i$ denotes the original model's output (or label), DiscoGP optimizes a composed objective that encourages three properties: (i) the discovered subnetwork should reproduce the original model's behavior on $\mathcal{D}$ (fidelity), (ii) the complement of the discovered circuit should perform poorly (completeness), and (iii) the circuit should be small (sparsity).

\begin{align*}
\mathcal{L}_{\text{fidelity}}(m)
&= -\sum_i \log p_{m}(\hat{y}_i \mid x_i)\\
\mathcal{L}_{\text{complete}}(m)
&= -\sum_i \mathbb{E}_K
[\log p_{\tilde{m}}(y_k \mid x_i)]\\
\mathcal{L}_{\text{sparse}}(m)
&= \mathbb{E}_{m_{E}} [\sigma(l_i)] + \mathbb{E}_{m_{\theta}} [\sigma(l_j)]
\end{align*}


Here, \(\tilde{m}=1-m\) denotes the complementary mask and \(K\), the number of labels; \(\mathcal{L}_{\text{complete}}\) therefore drives the complementary subnetwork toward near-uniform (random) predictions on \(\mathcal{D}\).  The per-coordinate \(\sigma(l_i)\) in \(\mathcal{L}_{\text{sparse}}\) serves as a continuous proxy for mask size.

The final training objective is a weighted sum:
$
\mathcal{L}_{\text{GP}}= \mathcal{L}_{\text{fidelity}} + \lambda_c\,\mathcal{L}_{\text{complete}} +\lambda_s\,\mathcal{L}_{\text{sparse}},
$
with hyperparameters \(\lambda_c,\lambda_s\) controlling the trade-offs between faithfulness, completeness, and compactness.

\begin{table*}[!htb]
\vspace{0.1in}
\noindent\makebox[\textwidth][l]{%
\resizebox{\textwidth}{!}{%
\begin{tabular}{@{} l | l | c c c c | c c c c @{}}
\hline
\multirow{2}{*}{\textbf{Classification Task}} & \multirow{2}{*}{\textbf{Setup}} 
  & \multicolumn{4}{c|}{\textbf{HuBERT}} & \multicolumn{4}{c}{\textbf{Wav2Vec 2.0}} \\ \cline{3-10}
 & & \textbf{Train Acc.} & \textbf{Test Acc.} & \textbf{Weight Circuit} & \textbf{Edge Circuit}
 & \textbf{Train Acc.} & \textbf{Test Acc.} & \textbf{Weight Circuit} & \textbf{Edge Circuit} \\ \hline

\multicolumn{10}{c}{\textbf{Articulatory Index}} \\ \hline
\multirow{8}{*}{Voicing Classification}  & Frozen    & 94.9\%  & 94.1\% & 100\% & 100\% & 84\% & 84.3\% & 100\% & 100\% \\ 
& Finetuned    & 98.1\%  & 96.9\% & 100\% & 100\% & 95.5\% & 94.4\% & 100\% & 100\% \\ \cline{2-10}
& DiscoGP  ($\mathcal{O}(H^2B^2)$)    & 98.1\%  & 96.9\% & 1.7\% & 0.8\% & 99.2\% & 97.7\% & 1.6\% & 0.3\% \\ 
& DiscoGP-W    & 98.8\%  & 97.2\% & 1.7\% & 100\% & 99.4\% & 97.9\% & 1.6\% & 100\% \\ 
& Random Head    & 97.6\% & 96.8\% & 1.2\% & 0.6\% & 99.4\% & 97.6\% & 1.3\% & 0.3\% \\ 
& Random Baseline    & 86.4\%  & 86.5\% & 1.7\% & 0.4\% & 89.6\% & 90.1\% & 1.2\% & 0.3\% \\ 
 \cline{2-10}
& QK DiscoGP (1/3)    & 98.7\%  & 97.1\% & 1.7\% & 2\% & 99\% & 97.2\% & 1.6\% & 0.5\% \\ 
& QKV ($\mathcal{O}(H^2B + HB^2)$)    & 98.8\%  & 97.4\% & 1.7\% & 2.7\% & 97.2\%  & 96.1\% & 1.6\% & 0.8\% \\ 
& QKV-MLP ($\mathcal{O}(HB)$)    & 95.2\%  & 94.6\% & 1.7\% & 1.7\% & 91.4\% & 91.3\% & 1.6\% & 0.4\% \\ 
 \hline

\multirow{8}{*}{Vowel Classification}  & Frozen    & 86.4\%  & 79.8\% & 100\% & 100\% & 57.6\% & 55.3\% & 100\% & 100\% \\ 
& Finetuned    & 91.4\%  & 82.5\% & 100\% & 100\% & 81\% & 77.8\% & 100\% & 100\% \\ \cline{2-10}
& DiscoGP  ($\mathcal{O}(H^2B^2)$)    & 91.8\%  & 89.2\% & 7.6\% & 3.3\% & 87.6\% & 85.6\% & 3.1\% & 0.6\% \\ 
& DiscoGP-W     & 87.7\%  & 84.9\% & 6.1\% & 100\% & 90.1\% & 88.4\% & 3.1\% & 100\% \\ 
& Random Head    & 91.1\%  & 88\% & 6.1\% & 1.8\% & 87.3\% & 85.3\% & 2.4\% & 2.7\% \\ 
& Random Baseline    & 76.7\%  & 77.3\% & 7.9\% & 1.1\% & 75\% & 74.3\% & 3.8\% & 0.4\% \\ 
 \cline{2-10}
& QK DiscoGP (1/3)    & 87.1\%  & 84.4\% & 7.6\% & 4.2\% & 91.2\% & 88.8\% & 3.1\% & 0.5\% \\ 
& QKV ($\mathcal{O}(H^2B + HB^2)$)    & 87.2\%  & 84.5\% & 7.6\% & 9.5\% & 91.1\%  & 90\% & 3.1\% & 0.8\% \\ 
& QKV-MLP ($\mathcal{O}(HB)$)    & 43.8\%  & 42.2\% & 7.6\% & 8.4\% & 53.9\% & 53.9\% & 3.1\% & 1.2\% \\ 
 \hline
\multirow{8}{*}{Consonant Classification}  & Frozen    & 85.7\% & 78.2\% & 100\% & 100\% & 51.2\% & 48.1\% & 100\% & 100\% \\ 
& Finetuned    & 92.4\%  & 85.5\% & 100\% & 100\% & 82.7\% & 75.7\% & 100\% & 100\% \\ \cline{2-10}
& DiscoGP  ($\mathcal{O}(H^2B^2)$)    & 84.5\% & 79.4\% & 6.5\% & 7\% & 85.5\% & 81.2\% & 3.9\% & 0.9\% \\ 
& DiscoGP-W    & 87.2\%  & 83.4\% & 6.5\% & 100\% & 85\% & 80.8\% & 3.9\% & 100\% \\ 
& Random Head    & 89.1\%  & 85\% & 4.3\% & 2.8\% & 84.2\% & 81.4\% & 2.9\% & 1\% \\ 
& Random Baseline    & 60.3\%  & 57.4\% & 4.8\% & 0.7\% & 60.7\% & 57.6\% & 5\% & 0.8\% \\ 
 \cline{2-10}
& QK DiscoGP (1/3)    & 83.5\% & 75.1\% & 6.5\% & 4.1\% & 87.3\% & 82.5\% & 3.9\% & 1.2\% \\
& QKV ($\mathcal{O}(H^2B + HB^2)$)    & 75.9\%  & 71.3\% & 6.5\% & 6.2\% & 87.4\%  & 83.6\% & 3.9\% & 1.5\% \\ 
& QKV-MLP ($\mathcal{O}(HB)$)    & 8.8\%  & 8.3\% & 6.5\% & 4.2\% & 37.1\% & 35.7\% & 3.9\% & 1.7\% \\ 
 \hline

\multicolumn{10}{c}{\textbf{Speech Commands dataset v1.0}} \\ \hline
\multirow{8}{*}{Keyword Spotting} & Frozen    & 98.1\%  & 97.8\% & 100\% & 100\% & 91.6\% & 89.8\% & 100\% & 100\% \\ 
& Finetuned    & 98.7\%  & 98.2\% & 100\% & 100\% & 98\% & 97.6\% & 100\% & 100\% \\ \cline{2-10}
& DiscoGP  ($\mathcal{O}(H^2B^2)$)    & 93.7\%  & 91.4\% & 5\% & 3.1\% & 95.9\% & 94.1\% & 2.7\% & 1.6\% \\ 
& DiscoGP-W   & 91.3\%  & 89.2\% & 5\% & 100\% & 95.6\% & 94.2\% & 2.7\% & 100\% \\ 
& Random Head    & 96.7\%  & 93.9\% & 2.7\% & 1.3\% & 96.2\% & 94.1\% & 1.4\% & 0.6\% \\ 
& Random Baseline    & 80.6\%  & 79.7\% & 4.6\% & 1.7\% & 79.1\% & 78.1\% & 2.8\% & 1.7\% \\ 
 \cline{2-10}
& QK DiscoGP (1/3)    & 92.8\%  & 90.5\% & 5\% & 4.2\% & 95.6\% & 93.7\% & 2.7\% & 2.3\% \\
& QKV ($\mathcal{O}(H^2B + HB^2)$)    & 93.6\%  & 91.1\% & 5\% & 3.9\% & 95.8\%  & 93.6\% & 2.7\% & 2\% \\ 
& QKV-MLP ($\mathcal{O}(HB)$)    & 71.7\%  & 70.1\% & 5\% & 4.9\% & 75\% & 73.4\% & 2.7\% & 2.6\% \\ 
 \hline

\multicolumn{10}{c}{\textbf{IEMOCAP}} \\ \hline
\multirow{8}{*}{Emotion Recognition} & Frozen    & 66\% & 63.4\% & 100\% & 100\% & 60.7\% & 59.6\% & 100\% & 100\% \\ 
& Finetuned    & 79.5\% & 70.7\% & 100\% & 100\% & 66\% & 61.7\% & 100\% & 100\% \\ \cline{2-10}
& DiscoGP  ($\mathcal{O}(H^2B^2)$)    & 62.8\% & 63.5\% & 1.9\% & 1.7\% & 64.9\% & 64\% & 1.2\% & 1.1\% \\ 
& DiscoGP-W  & 64.3\%  & 62.8\% & 1.7\% & 100\% & 63.7\% & 63\% & 1.2\% & 100\% \\ 
& Random Head    & 64.7\% & 63.5\% & 1.7\% & 2.2\% & 67.4\% & 65.3\% & 0.8\% & 2.1\% \\ 
& Random Baseline    & 59.2\% & 58.4\% & 2\% & 0.7\% & 58.8\% & 58.5\% & 1.4\% & 0.7\% \\ 
\cline{2-10}
& QK DiscoGP (1/3)    & 61\% & 60.8\% & 1.9\% & 2.6\% & 64.9\% & 64.5\% & 1.2\% & 1.6\% \\ 
& QKV ($\mathcal{O}(H^2B + HB^2)$)    & 60.3\%  & 60.7\% & 1.9\% & 3.3\% & 64.6\%  & 64.8\% & 1.2\% & 3.5\% \\ 
& QKV-MLP ($\mathcal{O}(HB)$)    & 58.1\%  & 59\% & 1.9\% & 3.2\% & 59.8\% & 59.9\% & 1.2\% & 3\% \\ 
 \hline

\multicolumn{10}{c}{\textbf{ASVspoof 2019}} \\ \hline
\multirow{8}{*}{Spoof Detection} & Frozen    & 97.6\%  & 95\% & 100\% & 100\% & 94.3\% & 92\% & 100\% & 100\% \\ 
& Finetuned    & 99.9\%  & 99.6\% & 100\% & 100\% & 99.9\% & 99.6\% & 100\% & 100\% \\ \cline{2-10}
& DiscoGP ($\mathcal{O}(H^2B^2)$)   & 95.1\%  & 96\% & 0.6\% & 2.4\% & 90\% & 88.7\% & 1.4\% & 1.3\% \\ 
& DiscoGP-W    & 96.1\%  & 93\% & 0.6\% & 100\% & 90.9\% & 89\% & 1.4\% & 100\% \\ 
& Random Head    & 92.4\%  & 93\% & 0.9\% & 0.6\% & 90.6\% & 90\% & 0.4\% & 2.3\% \\ 
& Random Baseline    & 90\%  & 89\% & 0.3\% & 1.5\% & 90\% & 89\% & 0.5\% & 1.6\% \\ 
 \cline{2-10}
& QK DiscoGP (1/3)    & 97.6\%  & 96\% & 0.6\% & 3.2\% & 90.8\% & 90\% & 1.4\% & 2.4\% \\
& QKV ($\mathcal{O}(H^2B + HB^2)$)    & 94.7\%  & 95.8\% & 0.6\% & 1.4\% & 97.8\%  & 95\% & 1.4\% & 1.6\% \\ 
& QKV-MLP ($\mathcal{O}(HB)$)    & 89.7\%  & 89.7\% & 0.6\% & 1\% & 93.8\% & 92\% & 1.4\% & 2\% \\ 
 \hline
\end{tabular}
} 
} 
\vspace{0.1in}
\caption{Performance and task competent circuit densities of HuBERT and Wav2Vec 2.0. We also show the result and corresponding asymptotic memory occupancies of circuit discovery with corresponding residual connections cut.}
\label{table:result}
\end{table*}
\section{Experimental Setup}
\label{exp}
\textbf{Datasets and tasks.}
We evaluate DiscoGP on several speech-encoding tasks using four datasets. For each experiment, we run three times and take the average except for emotion recognition where we perform 5-fold cross validation as there are no splits given. The \textbf{Articulatory Index} corpus~\cite{schatz2015articulation},  originally collected by ~\cite{fousek2004new}, contains pronunciations of CV and VC monosyllables from 20 English speakers (12 male, 8 female). For this dataset, we perform the following classification tasks:
\begin{itemize}[itemsep=0em]
  \item \textbf{Vowel classification:} classify the nucleus (15 vowels)
  \item \textbf{Consonant classification:} classify the onset/offset (24 consonants)
  \item \textbf{Voicing classification:}  onset/offset is voiced or unvoiced
\end{itemize}
To evaluate DiscoGP on longer inputs and under real-world settings, we perform circuit discovery on a keyword spotting task (by classifying the pronounced word with one of 10 keywords or \textit{unknown}) with the widely used \textbf{Speech Commands dataset v1.0} dataset~\cite{warden2017speech}, an emotion recognition task (predicting one of the four emotions: neutral, happy, sad, or angry from the speech input) using the \textbf{IEMOCAP} dataset~\cite{busso2008iemocap}, and spoof detection using the \textbf{ASVspoof 2019} dataset~\cite{9358099}.

\textbf{Models and experimental setups.}
We apply DiscoGP to two widely used speech encoders: \textbf{HuBERT} and \textbf{Wav2Vec 2.0}. 
For each, we evaluate four DiscoGP circuit discovery setups, and
compare two baselines with no weight or edge circuits discovered:

\begin{itemize}[itemsep=0em]
  \item \textbf{DiscoGP:} the base encoder is frozen and a one-layer feedforward network (FFN) is trained as the classification head
  \item \textbf{DiscoGP-W:} the first experiment with discovering only weight circuit (no edge circuit)
  \item \textbf{Random head:} the classification head is randomly initialized and kept fixed while the base encoder is unchanged
  \item \textbf{Random Baseline:} both the base encoder and the classification head are randomly initialized and kept fixed (a random baseline).

  \item \textbf{Frozen:} the base encoder is frozen and a one-layer feedforward network (FFN) is trained as the classification head
  \item \textbf{Finetuned:} the top two layers of the base encoder are finetuned together with the classification head.
\end{itemize}
We further identified the quartic growth in runtime memory occupancy of DiscoGP as a major bottleneck in deploying DiscoGP on larger models, datasets or smaller GPUs. in  ~\cref{memory} we identify the redundancy in residual connections is the main culprit of this quartic growth, and proposed multiple ways to cut the residual connections. The accuracy-memory tradeoff is then reported in ~\cref{table:result}.

\section{Results}

~\cref{table:result} summarizes the results for HuBERT and Wav2Vec 2.0. For each dataset and task we run DiscoGP three times and report the average accuracy on the train and test splits, as well as the average final circuit density in terms of retained weights and edges. In what follows we highlight the key findings, analyze ablation results, and offer hypotheses for notable behaviors.

\subsection{Analysis and Discussion}

\subsubsection{DiscoGP learns sparse, task-relevant circuits}
Across datasets and tasks, DiscoGP isolates extremely compact subnetworks while retaining test accuracy comparable to the frozen baseline. This shows the method can identify a small, task-relevant computation subgraph and weight parameters inside speech foundation models.

In addition, circuits discovered in randomly initialized models perform above chance but typically underperform circuits discovered in pretrained models by a substantial margin (commonly $\sim$10\% absolute, up to $\sim$25\%). 
If circuit discovery were merely constructing task networks ``out of noise,'' then completely randomly initialized networks should be able to match the performance of DiscoGP circuits, but they do not. This demonstrates that pretraining of HuBERT or Wav2Vec 2.0 carries useful, recoverable circuits into the weights and computational graphs.  This shows meaningful circuits are present in the pretrained model, and the gap between completely randomly initialized networks and DiscoGP quantifies how much of task performance is attributable to learned weight structure (not just architecture or mask selection). 

\subsubsection{Ablation study: Role of the classification head}
We replaced the learned classification head by a fixed random linear projection, since previous work using DiscoGP did not make use of encoder-only models and thus did not need task heads. This yields similar train and test accuracy, suggesting that the task-relevant computation is encoded in the foundation model activations and that circuit selection is representation- and encoding-driven rather than an artifact of a specific trained head.

\subsubsection {DiscoGP equalizes model performance gaps}
HuBERT generally gives stronger frozen baselines than Wav2Vec 2.0. Nevertheless, DiscoGP yields larger relative improvements for Wav2Vec 2.0. Masking makes specialized circuits emerge from its more general parameters. Discovered circuits for Wav2Vec 2.0 give comparable performance to frozen HuBERT and generally far exceed frozen Wav2Vec 2.0.

\subsubsection{DiscoGP can improve generalization}
The discovered circuits mostly outperform the frozen baseline on test sets, and often maintain their high performance towards the end. Two mechanisms may explain this: (1) \emph{extra parameters and training}: circuit discovery introduces additional trainable variables (mask logits for weights and edges) and typically runs extra optimization steps, which provide a constrained but nontrivial adaptation capacity similar to finetuning; (2) \emph{implicit regularization}: binary masking removes parameters that may overfit or carry noisy signals (we observe lower train but higher test accuracy in many cases). Finetuning usually achieves the highest absolute performance as it is not meant to find sparse circuits and can therefore adjust every weight and connection freely.

\subsubsection{Weight vs.\ edge circuits}
The benefits of weight circuit only compared to the final circuit depend on the task:
For vowel classification, keyword spotting, and spoof detection on HuBERT and consonant classification and emotion recognition on Wav2Vec 2.0, edge circuit improves accuracy. For the rest of the tasks, edge circuit reduces performance somewhat. Together, these observations suggest that DiscoGP can benefit from an early stopping mechanism during edge circuit discovery.

\section{Memory-Efficient DiscoGP}
\label{memory}
\begin{figure*}[t!]
    \begin{subfigure}[]{0.5\textwidth}
    \centering
    \includegraphics[width=\linewidth]{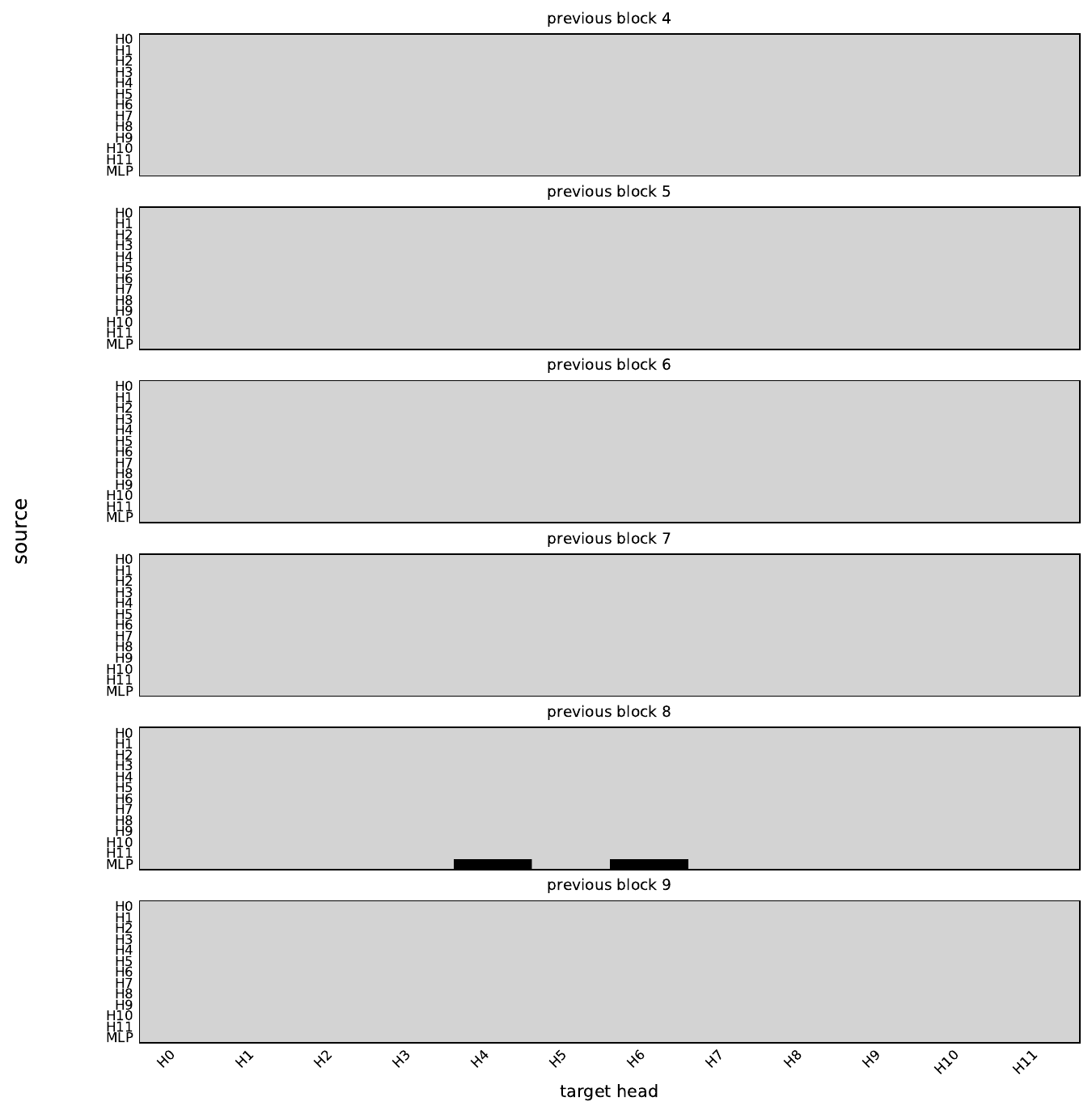}
    \caption{Block 10's K edge mask}
    \label{fig:figure1b}
    \end{subfigure}
    \begin{subfigure}[]{0.5\textwidth}
    \includegraphics[width=\linewidth]{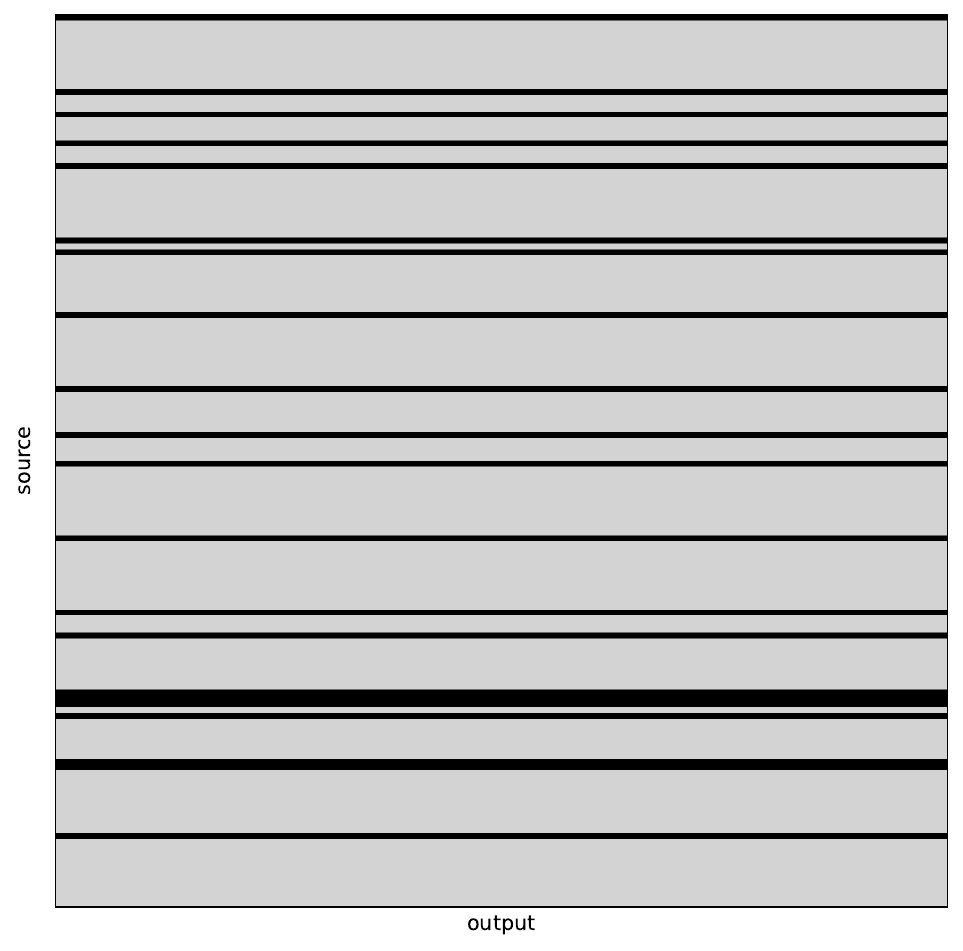}
    \caption{Block 10's MLP edge mask}
    \label{fig:figure1c}
    \end{subfigure}
    \caption{(a) Block 10's K binary edge masks of the vowel classification task with residuals coming form block 4 onward. light greys are 0 and blacks are 1. It is clear that most of the previous residuals are turned off by edge masks anyway. (b) Block 10's K binary edge masks of the vowel classification task with all previous residuals. light greys are 0 and blacks are 1. There is no clear pattern as to which residuals are removable.}
    \label{fig:figure1}
    \vspace{-0.2in}
\end{figure*}
A major runtime bottleneck in DiscoGP is the memory required to discover edge circuits, which grows quartically with model depth and can easily exceed the capacity of a single GPU. For example, models such as HuBERT and Llama3.2-1B cannot be hosted on an NVIDIA L4 (24\,GB) once edge-circuit discovery is enabled. For those that can be hosted, the batch size sometimes has to be reduced to 1 or 2. We therefore modify DiscoGP to reduce the memory footprint of edge-circuit discovery from asymptotically quartic to asymptotically cubic.

\subsection{Memory-Efficient Edge Circuit Discovery}
Let $B$ denote the number of blocks and $H$ the number of attention heads per block. In the original formulation, the number of edge-mask parameters


\begin{equation}
\begin{aligned}
   &= |\text{Output Edge Circuit}|\\
   &\quad {}+\sum_{i=0}^{B-1} \Bigl(|\text{QKV Edge Circuit}| + |\text{MLP Edge Circuit}|\Bigr) \\
&= [(H+1)B + 1]\\
   &\quad {}+\sum_{i=0}^{B-1} \left\{3[(H+1)i+1]H + [(H+1)i+1+H]\right\} \\
    &\approx \mathcal{O}\!\left(H^2B^2\right).
\end{aligned}
\label{eq:total_complexity}
\end{equation}

The dominant term arises from residual-stream nodes accumulated across blocks: as depth increases, every new block can connect to an ever-growing set of earlier activations. We observe that the learned \textbf{Q} and \textbf{K} masks zeroed most early residual nodes. Based on this observation, we propose to keep only the last $k$ blocks in the edge masks and zero out residual nodes from earlier blocks. In practice, we set $k=2$, retaining only the current and previous block. We empirically verify our observation by evaluating the original setup and three new method of DiscoGP with QK residual removed, QKV residual removed, and QKV+MLP residual removed on GPT2, Llama3.2-1B, HuBERT, and Wav2Vec 2.0 models and their corresponding text and speech tasks with original DiscoGP running on NVIDIA L40 (48GB) and NVIDIA H100 (80GB) GPUs, and the new DiscoGP entirely on NVIDIA L40 GPUs.

The results in~\cref{table:result} show the accuracy-memory tradeoff in detail across all models and tasks we considered. Overall, the results shows that the QKV residuals in edge-circuit discovery can be safely removed without largely impairing accuracy, but the MLP residuals must remain intact. This is reasonable because attention heads mainly act as routing machinery instead of information storage. Prior works~\cite{NEURIPS2019_2c601ad9, clark2019what} have shown that many attention heads can be pruned while retaining performance, which suggests substantial redundancy in the QKV pathway. In contrast, Geva et al.~\cite{geva-etal-2021-transformer} point out that MLP residual are the information storage, where each layer's output continuously retain, modify or veto.

In our new DiscoGP algorithm with QKV residuals zeroed out, the resulting parameter count 
\begin{equation}
\begin{aligned}
&= |\text{Output Edge Circuit}|\\ 
&\quad {}+ \sum_{i=0}^{B-1} \Bigl(|\text{QKV Edge Circuit}| + |\text{MLP Edge Circuit}|\Bigr) \\
&=[(H + 1)B + 1]\\ 
&\quad {}+ \sum_{i = 0}^{B - 1} \{3kH(H+1) + [(H+1)i+1+H]\}\\
    &\approx \mathcal{O}(H^2B + HB^2)
\end{aligned}
\label{eq:cut_complexity}
\end{equation}

Similarly, the runtime memory cost of other residual cutting paradigms can be calculated, and the trade off between memory occupation and task performances are outlined in~\cref{table:result}. Based on the results, in terms of the number of blocks and heads of a model, our method with QKV residuals zeroed attains the optimal trade off, which reduces the memory required for edge-circuit discovery from quartic to cubic while mostly maintains the performance.

We also verify the generalization of our residual cut but evaluating our algorithms on the text models (GPT2~\cite{radford2019gpt2} and Llama3.1-1B Instruct~\cite{grattafiori2024llama3herdmodels}) with Blimp~\cite{warstadt2020blimp}, IOI~\cite{wang2022interpretability}, and Pararel~\cite{elazar2021measuring} tasks defined in the original DiscoGP paper~\cite{yu-etal-2025-sheaf}. The results show that QKV residuals zeroed attains the optimal trade off as well. The results are reported in the appendix.

\section{Discussion}
Our results suggest three broader implications of circuit discovery for speech encoders. First, circuit discovery isolates compact \emph{computational subgraphs} (often retaining single-digit percentages of weights/edges) that reproduce the original encoder + head performance. The resulting sparsity is best viewed as evidence that the task can be explained by a small subset of the original computation, rather than as a compression objective in its own right. As a practical side effect, these sparse circuits are also easier to store and run, which have to potential for an alternative lower-memory and lower-latency deployment.

Second, circuit discovery yields a more \emph{causal} interpretability signal than layerwise diagnostic probes. Probes can produce false positives by decoding features that are present but not used, whereas, by optimizing sparse binary masks over both weights and computation-graph edges, DiscoGP explicitly selects the chain of operations required to produce the output, yielding a direct test of whether pretrained representations are \emph{used}.

Third, the circuit view moves analysis away from monosemantic single-neuron explanations toward \emph{functional groups} of units operating in superposition. By jointly discovering weights circuits and unrolled residual-stream edges, DiscoGP uncovers small, interacting subnetworks whose collective computation implements the task.

In a separate contribution, we also develop a memory-efficient variant of DiscoGP that reduces GPU memory usage during edge-circuit discovery from quartic to cubic in terms of the number of blocks and heads of the model. This does not change the discovered circuits themselves; rather, it makes the discovery procedure more practical on limited hardware and less likely to trigger CUDA out-of-memory errors.

\section{Limitations and future work}

Our evaluation is limited to two encoder families (HuBERT and Wav2Vec~2.0) and English datasets. Extending the method to other architectures, pretraining objectives, and typologically diverse languages may improve generality.

Our experiments were also restricted to classification tasks. While we aim to extend this to regression tasks, doing so presents specific challenges. Validating circuit discovery on regression tasks remains to be done in both text-based language models and speech models, and so we currently lack a robust framework demonstrating this.

Finally, we remark that we observed a consistent phenomenon whereby the weight density of the discovered circuits decreases from earlier layers to later layers across all evaluated tasks. We only document this trend here; a full theoretical understanding and causal explanation of this progression of sparsity require further experimental investigation.

\section{Summary of Contributions}
We extend Mechnistic Interpretation to modern speech encoders, including HuBERT and Wav2Vec 2.0, and show that it uncovers compact, task-specific circuits across a range of speech tasks. We provide empirical validation and targeted ablations showing that these circuits reflect pretrained computation rather than random structure or task-head artifacts, and we analyze several plausible explanations for the observed generalization gains. Separately, we introduce a memory-efficient DiscoGP variant that reduces GPU memory usage during circuit discovery, making the method more practical on limited hardware. Together, these results broaden mechanistic interpretability beyond text decoders and show that circuit-level analysis can support both explanatory insight and practical use.

\section{AI-Generated Content Disclosure}
We used ChatGPT to review grammar, improve clarity of expression, and refine wording in this manuscript. We did not use generative AI tools to draft substantive portions of the paper or to assist with the conceptualization of the research, methodological design, experimental execution, data analysis, or interpretation of the results. All scientific contributions, decisions, and conclusions presented in this work are solely our own.

\FloatBarrier
\bibliographystyle{IEEEtran}
\bibliography{IEEEexample}
\end{document}